\documentclass{article}

\pdfoutput=1

\usepackage{times}
\usepackage{graphicx} %
\usepackage{subfigure} 

\usepackage{natbib}

\usepackage{algorithm}
\usepackage{algorithmic}

\usepackage{amsfonts}

\usepackage{amsmath}

\usepackage[nohyperref,accepted]{icml2015}

\newcommand{\argmin}[1]{\underset{#1}{\mathrm{argmin}}}

\newcommand{\km}{$k$-mer}
\newcommand{\la}{\ensuremath{\leftarrow}}

\newcommand{\phib}{\ensuremath{{\pmb \phi}}}

\newcommand{\istar}{\ensuremath{{i^\star}}}

\newcommand{\Rstar}{\ensuremath{{\mathcal{R}^\star}}}

\newcommand{\Ncal}{\ensuremath{\mathcal{N}}}
\newcommand{\Pcal}{\ensuremath{\mathcal{P}}}
\newcommand{\Rcal}{\ensuremath{\mathcal{R}}}
\newcommand{\Scal}{\ensuremath{\mathcal{S}}}

\newcommand{\xb}{\ensuremath{\mathbf{x}}}
\newcommand{\eqdef}{\overset{{\rm \mbox{\tiny def}}}{=}}

\newlength\myindent
\icmltitlerunning{Greedy Biomarker Discovery in the Genome with Applications to Antimicrobial Resistance}

\begin{document} 

\twocolumn[
\icmltitle{Greedy Biomarker Discovery in the Genome with Applications to Antimicrobial Resistance}

\icmlauthor{Alexandre Drouin$^1$}{alexandre.drouin.8@ulaval.ca}
\icmlauthor{S\'{e}bastien Gigu\`{e}re$^3$}{giguere.sebastien@gmail.com}
\icmlauthor{Maxime D\'{e}raspe$^2$}{maxime.deraspe.1@ulaval.ca}
\icmlauthor{Fran\c{c}ois Laviolette$^1$}{francois.laviolette@ift.ulaval.ca}
\icmlauthor{Mario Marchand$^1$}{mario.marchand@ift.ulaval.ca}
\icmlauthor{Jacques Corbeil$^2$}{jacques.corbeil@genome.ulaval.ca}
\icmladdress{$^1$ Department of Computer Science and Software Engineering, $^2$ Department of Molecular Medicine, Laval University, Quebec, Canada; $^3$ Institute for Research in Immunology and Cancer, University of Montreal, Montreal, Canada}

\icmlkeywords{set covering machine, genomics, sparse, greedy}

\vskip 0.3in
]

\begin{abstract}
The Set Covering Machine (SCM) is a greedy learning algorithm that produces sparse classifiers.
We extend the SCM for datasets that contain a huge number of features.
The whole genetic material of living organisms is an example of such a case, where the number of feature exceeds $10^7$.
Three human pathogens were used to evaluate the performance of the SCM at predicting antimicrobial resistance.
Our results show that the SCM compares favorably in terms of sparsity and accuracy against $L_1$ and $L_2$ regularized Support Vector Machines and CART decision trees.
Moreover, the SCM was the only algorithm that could consider the full feature space.
For all other algorithms, the latter had to be filtered as a preprocessing step.
\end{abstract} 

\section{Introduction}

Genomics is a discipline of biology that focuses on analysing the entire genetic material of individuals, which is called the genome.
Recent advances in next-generation sequencing (NGS) have led to a tremendous increase in the affordability of whole genome sequencing~\cite{vandijk2014}.
The reduced cost and increased throughput of NGS have motivated its use for case-control studies, where groups of individuals are compared based on their genomes~\cite{hall2013, vandijk2014}.
Such studies can serve to determine the genomic variations that are biomarkers (i.e.: measurable characteristics) of a given biological state (phenotype).
Identifying such biomarkers has important implications in the clinical setting, where they can serve as the basis for diagnostic tests.
Moreover, they can guide the development of new personalised therapies or drug treatments, by providing insight on the biological processes that are responsible for a phenotype.

It is common to represent a genome by a set of single nucleotide polymorphisms (SNP) ~\cite{brookes1999}.
A SNP exists at a single base pair location in the genome when a variation occurs within a population.
They are obtained by aligning multiple genomes, a computationally expensive task that can be affected by gene deletions, duplications, inversions, or translocations~\cite{leimeister2014}.
To address these limitations, we favor an approach, inspired by the ``bag-of-words'' representation, that is heavily used in the domain of text classification and string kernels.
It consists in representing each genome by all its constituent \km s, i.e. all the substrings of length $k$ that are contained in the genome.

In the context of biomarker discovery, one is interested in finding the smallest subset of genomic features that allows to accurately predict the phenotype.
Including superfluous features in this subset, would lead to the development of unnecessarily complicated diagnostic tests, generating additional costs.
This is a challenging machine learning problem on many aspects.
First, only a small fraction of the \km~features are likely to be associated with the phenotype.
Second, some \km s are naturally highly correlated, as they belong to the same gene or gene family.
Third, for genomes, the number of learning examples is often much smaller than the total number of possible \km s.
Therefore, one must use a method that favors sparsity and that is able to retrieve important features from such an extremely high dimensional feature space, while at the same time avoiding overfitting.

In this paper, we propose a method for learning sparse and interpretable models from whole genomes for predicting discrete phenotypes.
Our approach relies on the Set Covering Machine~\cite{marchand2003}, a greedy learning algorithm that produces highly sparse models and that achieved state-of-the-art accuracy for many learning tasks, such as learning from DNA microarray data~\cite{shah2012}. The obtained models are short conjunctions or disjunctions of boolean-valued rules, which can explicitly highlight the importance of specific DNA sequences.

The next section presents the Set Covering Machine algorithm together with some improvements.
Then, we explain how the SCM can be used to learn from genomes.
Finally, the algorithm is used to predict the antimicrobial resistance of three common human pathogens for a panel of antibiotics.
The results are then compared to the ones of $L_1$ and $L_2$ regularized Support Vector Machines~\cite{cortes1995} and CART decision trees~\cite{breiman1984} based on risk and sparsity.

\section{Methods}

\subsection{The Set Covering Machine}

In the supervised machine learning setting, we assume that data are available as a set $\Scal=\{(\xb_i,y_i)\}_{i=1}^m \sim D^m$, where $\xb_i \in \mathcal{X}$ is a training example, $y_i \in \mathcal{Y}$ its associated label and $D$ is an unknown data generating distribution.
We consider binary classification problems where $\mathcal{Y} = \{0, 1\}$. 
The goal of a learning algorithm is to produce a predictor $h: \mathcal{X} \rightarrow \mathcal{Y}$ that minimizes the expected risk, which is given by:
\begin{equation}
\underset{(\xb, y) \sim D}{E} I[h(\xb) \not = y],
\end{equation}
with $I[True] = 1$ and 0 otherwise.

The Set Covering Machine (SCM) \citep{marchand2003}, is a learning algorithm that produces predictors that are conjunctions or disjunctions of boolean-valued rules $r : \mathcal{X} \rightarrow \{0, 1\}$. 
Given a set of rules $\Rcal$, the SCM attempts to find a predictor that minimizes the empirical risk $R_\mathcal{S} \eqdef \sum_{i=1}^m I[h(\xb_i) \not = y_i] / m$, while using the smallest subset of $\Rcal$.
This problem is reducible to the minimum set cover problem, which is known to be $NP$-hard \citep{haussler1988, marchand2003}.
To overcome this issue, the SCM uses a greedy optimisation algorithm inspired by the algorithm of \citet{chvatal1979}, which yields an approximate solution with a worst case guarantee.
In addition, \citet{germain2012pseudo} used combinatorial optimisation to show that the solution found using the greedy heuristic is very close to optimality in most cases.
Algorithm \ref{algo:scm_conj} presents the SCM algorithm for the case where the returned predictor is a conjunction of boolean-valued rules.
For the sake of conciseness, we only present the conjunction case.
The disjunction case can be obtained from the previous one by using $\Scal'=\{(\xb_i, \neg y_i) : (\xb_i, y_i) \in \Scal\}$ as the set of training examples and taking the complement of the returned predictor $h$. 
This follows from the De Morgan law: $\neg (\bigwedge_{r^\star \in \Rstar} r^\star(\xb)) = \bigvee_{r^\star \in \Rstar} \neg r^\star(\xb) $.%
The SCM starts with an empty conjunction and extends it in a greedy manner by iteratively selecting the rule maximizing a utility function.
The latter is designed to favor conjunctions that correctly classify most of the training examples, while taking into account some constraints imposed by a greedy approach.
Indeed, since the algorithm is greedy, once a rule is added to the conjunction, it cannot be removed.
Observe that, any rule in the conjunction that assigns the negative class to an example forces the result of the conjunction itself to be negative.
Therefore, there are two types of errors to consider: making an error on a negative example, which can be recovered, and making an error on a positive example, which cannot be recovered.
For this reason, \citet{marchand2003} propose to score each rule using the following utility function:
\begin{equation}
U \eqdef |\mathcal{A}| - p \cdot |\mathcal{B}|,
\end{equation}
where $|\mathcal{A}|$ is the number of negative examples that it correctly classifies (i.e., covered by that rule) and $|\mathcal{B}|$ is the number of positive examples on which it errs.
A hyperparameter $p$, that is usually selected by cross-validation, allows to fix the correct trade-off between these two types of errors.

This process is repeated until a stopping criterion is reached.
However, at each iteration the examples that are classified as negative by the selected rule are discarded for further computations of the utility function.
This is justified by the observation above, that is, for those examples, the result of the conjunction is necessarily negative.
This ensures that redundant rules are not added to the conjunction, effectively favoring sparse models.

There are 3 stopping criteria.
The first stopping criterion is reached when all the negative examples have been covered by the rules of the conjunction.
In this case, there is no need to continue extending the conjunction, as it is consistent with all the negative examples and adding more rules can only lead to more errors on the remaining positive examples. 
The second stopping criterion is reached when the number of rules in the conjunction reaches the limit~$s$.
This limit is a hyperparameter that induces regularization by early-stopping.
Finally, the third stopping criterion is reached when the rule of maximal utility is consistent with no negative examples and errs on no positive examples. 
In this case, the algorithm is in a state of equilibrium, since no examples are removed at the end of the iteration.

Note that the version of the SCM presented here differs in two points from the one of \citet{marchand2003}.
First, when more than one rule have the maximal utility, it selects the rule with the smallest empirical risk.
This simple strategy is important for genomic datasets where the number of features is much greater than the number of examples.
It becomes particularly important after a few iterations, as fewer examples contribute to the utility function and a lot of rules may have the same utility.
In this situation, it is reasonable to assume that the rule that has the best performance on all the examples of the training set, is more likely to contribute to the best generalization performance.
Second, the algorithm is stopped when it reaches the state of equilibrium mentioned above.
This prevents from adding useless rules and reduces the training time.

The worst-case running time complexity of Algorithm \ref{algo:scm_conj} is $O(|\mathcal{R}| \cdot |\Scal| \cdot s)$. 
It thus scales linearly in the number of rules and the number of training examples.

\begin{algorithm}[tb]
   \caption{Set Covering Machine (Conjunction)}
   \label{algo:scm_conj}
\begin{algorithmic}
   \STATE {\bfseries Input:} $\Scal$: A set of $m$ training examples, $\Rcal$: A set of boolean-valued rules, $p$: The trade-off parameter, $s$: The early stopping parameter
   
   \STATE $\Rstar \la \emptyset$
   \STATE $\Pcal \la \mbox{the set of examples in $\Scal$ with label $1$}$
   \STATE $\Ncal \la \mbox{the set of examples in $\Scal$ with label $0$}$
   \STATE $stop \la False$

   \WHILE{$\mathcal{N} \not= \emptyset$ \AND $|\Rstar| < s$ \AND $\neg stop$} 
   
   \STATE $\triangleright$ Compute the utility function for each rule
   \STATE ~~~$\forall i \in \{ 1, ..., |\mathcal{R}| \}$,
   \STATE ~~~ ~~~~~$\mathcal{A}_i \la$ the subset of $\mathcal{N}$ correctly classified by $r_i$
   \STATE ~~~ ~~~~~$\mathcal{B}_i \la$ the subset of $\mathcal{P}$ misclassified by $r_i$
   \STATE ~~~ ~~~~~$U_i \la |\mathcal{A}_i| - p \cdot |\mathcal{B}_i|$

   \STATE $\triangleright$ Select the best rule
   \STATE ~~~$U^\star \la \underset{i \in \{ 1, ..., |\mathcal{R}| \}}{\max} U_i$
   \STATE ~~~$\mathcal{C} \la \{ i\in \{ 1, ..., |\mathcal{R}| \}~|~U_i = U^\star \}$
   \STATE ~~~$i^\star \la \argmin{i \in \mathcal{C}}~ \sum_{j=1}^m I[r_i(\xb_j) \not = y_j] / m$
  
   \STATE $\triangleright$ Add the best rule to the conjunction
   \STATE ~~~ {\upshape \textbf{if}} $|\mathcal{A}_\istar| > 0$ \OR $|\mathcal{B}_\istar| > 0$ {\upshape \textbf{then}}
   \STATE ~~~ ~~~~~$\Rstar \la \Rstar \cup \{r_{\istar}\}$
   \STATE ~~~ ~~~~~$\mathcal{N} \la \mathcal{N} - \mathcal{A}_{\istar}$
   \STATE ~~~ ~~~~~$\mathcal{P} \la \mathcal{P} - \mathcal{B}_{\istar}$
   \STATE ~~~ {\upshape \textbf{else}}
   \STATE ~~~ ~~~~~$stop \la True$
   \STATE ~~~ {\upshape \textbf{endif}}
   
   \ENDWHILE
   
   \STATE {\upshape \textbf{return}} $h$, where $h(\xb) = \bigwedge_{r^\star \in \Rstar} r^\star(\xb)$
   
\end{algorithmic}
\end{algorithm}

\subsection{Applying the Set Covering Machine to Genomes}

We represent each genome by the presence or absence of every possible \km.
Let $\mathcal{K}$ be the set of all, possibly overlapping, \km s present in at least one genome of the training set.
We can safely omit \km s that are not in $\mathcal{K}$ as they could not serve to discriminate genomes of the training set.
For each genome $\xb$,  we define a vector $\phib(\xb) \in \{0,1\}^{|\mathcal{K}|}$, such that $\phi_i(\xb) = 1$ if the \km~$k_i \in \mathcal{K}$ is in $\xb$ and $0$ otherwise.
We then define a new training set $\Scal'=\{(\phib(\xb_i), y_i) : (\xb_i, y_i) \in \Scal\}$.

The set of boolean-valued rules that we consider is composed of 2 types of rules: presence rules and absence rules, which rely on the $\phib(\xb)$ vectors to determine their outcome.
For each \km~$k_i \in \mathcal{K}$, we define a presence rule as $p_{k_i}(\phib(\xb)) \eqdef I[\phi_i(\xb) \!=\! 1]$ and an absence rule as $a_{k_i}(\phib(\xb)) \eqdef I[\phi_i(\xb) \! =\! 0]$.
The rules for each \km~in $\mathcal{K}$ are then combined to form the set $\Rcal$.

The SCM (Algorithm \ref{algo:scm_conj}), can then be applied with $\Scal'$ as the training set and $\Rcal$ as the set of boolean-valued rules.
This yields a predictor which explicitly highlights the importance of a small set of \km s for predicting the phenotype. 
In addition, this predictor has a form which is simple to interpret, since its predictions are the outcome of a simple logical operation.

\section{Results and Discussion}

\begin{table*}[htbp]
\caption{Results for the Set Covering Machine (SCM), the CART algorithm (CART), $L_1$/$L_2$ regularized Support Vector Machines (L1SVM, L2SVM) and the baseline (Dummy). The prefix $\chi^2$ indicates that a univariate filter was applied prior to learning. For each dataset, the values are the average risk and number of \km s in the model (in parenthesis) for the 5 folds. The best risks are in bold.}
\label{tbl:all_res}
\vskip 0.15in
\begin{center}
\begin{small}
\begin{sc}
\begin{tabular}{l|llllll}
\hline
\abovespace\belowspace
Dataset & SCM & $\chi^2 + $ SCM & $\chi^2 + $ CART & $\chi^2 + $ L1SVM & $\chi^2 + $ L2SVM & Dummy\\
\hline
\textbf{C. difficile} \\
\,\, Azithromycin & \textbf{0.015} (3.2) & 0.024 (4.8) & 0.035 (6.6)  & 0.020 (494.6) & 0.035 (2451870.2) & 0.461 \\
\,\, Ceftriaxone & \textbf{0.070} (2.0) & 0.130 (5.6) & 0.112 (7.2) & 0.091 (277.8) & 0.091 (2332313.0) & 0.305 \\
\,\, Clarithromycin & \textbf{0.015} (3.0) & 0.019 (4.6) & 0.026 (7.6) & 0.022 (522.6) & 0.041 (2426505.8) & 0.461 \\
\,\, Clindamycin & 0.025 (2.0) & 0.025 (2.4) & 0.008 (2.4) & \textbf{0.006} (702.2) & 0.03 (2405735.4) & 0.140 \\
\,\, Moxifloxacin & \textbf{0.019} (1.0) & 0.030 (1.8) & \textbf{0.019} (1.0) & 0.022 (173.6) & 0.048 (2432399.0) & 0.407 \\
\hline
\textbf{P. aeruginosa} \\
\,\, Amikacin & \textbf{0.181} (6.0) & 0.208 (9.8) & 0.211 (18.8) & 0.222 (687.8) & 0.186 (164778.2) & 0.230 \\
\,\, Doripenem & 0.234 (1.4) & 0.237 (1.6) & 0.242 (25.4) & \textbf{0.220} (44.8) & 0.237 (16614.2) & 0.377 \\
\,\, Meropenem & 0.280 (1.8) & 0.272 (1.8) & 0.283 (9.2) & \textbf{0.253} (233.6) & 0.256 (3475.6) & 0.416 \\
\,\, Levofloxacin & 0.067 (1.4) & \textbf{0.058} (1.8) & 0.067 (1.0) & 0.081 (180.4) & 0.137 (173177.4) & 0.472 \\
\hline
\textbf{S. pneumoniae} \\
\,\, Benzylpenicillin & \textbf{0.012} (1.0) & \textbf{0.012} (1.2) & 0.012 (1.8) & 0.019 (295.8) & 0.017 (550134.8) & 0.076 \\
\,\, Erythromrycin & \textbf{0.031} (2.0) & 0.047 (5.6) & 0.045 (4.4) & 0.034 (299.4) & 0.041 (476701.6) & 0.142 \\
\,\, Tetracyclin & \textbf{0.025} (1.2) & \textbf{0.025} (2.2) & 0.028 (1.0) & \textbf{0.025} (479.8) & \textbf{0.025} (516480.4) & 0.111 \\
\hline
\hline
Average & \textbf{0.081} (2.2) & 0.091 (3.6) & 0.091 (7.2) & 0.085 (366.0) & 0.095 (1162515.5) & 0.300 \\
\hline
\end{tabular}
\end{sc}
\end{small}
\end{center}
\vskip -0.1in
\end{table*}

We applied the SCM and our proposed data representation to a real-world biomarker discovery problem, which consists in predicting the antimicrobial resistance bacteria based on their genomes.
Antimicrobial resistance is a growing public health concern, as many multi-drug-resistant strains are starting to emerge.
This compromises our ability to treat common infections, which results in an increasing number of deaths and health care costs~\cite{who2014}.
An accurate predictor of antimicrobial resistance, could allow faster profiling of drug-resistant strains.

We present results for three human pathogens: \textit{Clostridium difficile}, \textit{Pseudomonas aeruginosa}~\cite{kos2015} and \textit{Streptococcus pneumoniae}~\cite{croucher2013}.
For each of the latter, $285$ to $556$ bacterial isolates were collected from patients across the world.
The genome of each isolate was sequenced and their susceptibility was measured against a panel of antibiotics.
We considered each (pathogen, antibiotic) combination individually, yielding $12$ datasets in which the number of \km s ($|\mathcal{K}|$) ranges from $10,542,251$ to $132,487,288$.
Note that we consider \km s of length $31$, as this value is often used for bacterial genome assembly.

We empirically compared the risk and sparsity of models obtained using the SCM, $L1$ and $L2$ regularized SVMs~\cite{cortes1995} and the CART decision tree algorithm~\cite{breiman1984}.
We used the SVM implementation from LIBLINEAR~\cite{fan2008} and the CART implementation from Scikit-learn~\cite{pedregosa2011}.

The great number of features that we consider poses computational challenges in terms of runtime and memory usage.
The simplicity of the SCM and its low computational complexity enabled us to implement the algorithm in a way that made it possible to learn from the entire feature space.
However, for SVM and CART, dimensionality reduction was necessary.
For these algorithms, we filtered the features using a univariate filter~\cite{guyon2003}, with the $\chi^2$ test as the measure of significance and the method of \citet{benjamini2001} for multiple testing correction.

Table \ref{tbl:all_res} presents the 5-fold nested cross-validation risk and the average number of \km s in the model for each dataset and learning algorithm.
For each fold, the hyperparameters were selected using standard cross-validation on the remaining 4 folds. This table also includes a comparison to a baseline (Dummy), that predicts the majority class in the training set.

Observe that the SCM tends to learn models that are much sparser than the ones of SVMs.
This is an interesting result, as both the SCM and the L1SVM algorithms attempt to obtain sparse solutions.
This suggests that the greedy heuristic of the SCM is much more efficient at minimizing the $L_0$ norm than the L1SVM.
Moreover, although the difference in sparsity is less striking, the SCM tends to learn sparser models than CART.

In addition, note that all the algorithms clearly outperform the dummy predictor, which means that some information on antimicrobial resistance is contained in the genomes.
It can also be observed that, for 8 of the 12 datasets, the risks of the SCM predictors are smaller or equal to the ones of the other algorithms.
This suggests that the extreme sparsity of the former does not undermine their generalization performance.

Finally, we compared the SCM to a variant which uses a univariate filter as a preprocessing step ($\chi^2 + $ SCM).
On some datasets, using such a filter leads to an increased risk and denser models.
Being able to consider the entire feature space without filtering is thus an interesting property of the SCM algorithm.

\section{Conclusion}

In this work, we have confronted the Set Covering Machine to the challenging problem of learning from extremely high dimensional feature spaces and obtaining sparse models.
The analysis was conducted in the context of biomarker discovery, which is a problem of high importance. %
We showed that, as opposed to other learning algorithms, the Set Covering Machine can learn from entire genome sequences without requiring prior feature selection.
Our results for predicting antimicrobial resistance suggest that the greedy heuristic of the SCM produces sparser models than a Support Vector Machine with a $L_1$ regularizer, while having similar and often better generalization performance.
To the best of our knowledge, this is the first time that Set Covering Machines are used on datasets of such high dimensionality.
The fact that the obtained models are sparse and generalize well, opens the door to new applications in other fields where datasets of high dimensionality are common, such as genome-wide association studies (GWAS) and natural language processing.

\section*{Acknowledgments}

We thank Dr Veronica Kos, Dr Humphrey Gardner and their colleagues from AstraZeneca for providing
the \textit{Pseudomonas aeruginosa} data. 
We also thank Dr Vivian Loo and Dr Anne-Marie Bourgault for sharing the \textit{Clostridium difficile} data. 
Computations were performed on the Colosse supercomputer at Universit\'{e} Laval (resource allocation project: nne-790-ae), under the auspices of Calcul Qu\'{e}bec and Compute Canada. 
AD is recipient of an Alexander Graham Bell Canada Graduate Scholarship Doctoral Award from the National Sciences and Engineering Research Council of Canada (NSERC). 
This work was supported in part by the Fonds de recherche du Qu\'{e}bec - Nature et technologies (FL, MM \& JC; 2013-PR-166708), the NSERC Discovery Grants (FL; 262067, MM; 122405) and an award to Michael Tyers from the Minist\`{e}re de l'enseignement sup\'{e}rieur, de la recherche, de la science et de la technologie du Qu\'{e}bec through G\'{e}nome Qu\'{e}bec (SG). 
JC acknowledges the Canada Research Chair in Medical Genomics.

\bibliography{example_paper}
\bibliographystyle{icml2015}

\end{document}